\documentclass[epj-spec]{svjour}
\usepackage{graphics}
\begin{document}
\title{Weak localization in monolayer and bilayer graphene}
\author{K. Kechedzhi\inst{1}\fnmsep\thanks{\email{k.kechedzhi@lancaster.ac.uk}}
\and E. McCann\inst{1} \and Vladimir I. Fal'ko\inst{1} \and H.
Suzuura\inst{2} \and T. Ando\inst{3} \and B. L. Altshuler\inst{4}}
\institute{Department of Physics, Lancaster University, Lancaster,
LA1 4YB,~UK \and Division of Applied Physics, Graduate School of
Engineering, Hokkaido University, Sapporo 060-8628, Japan \and
Department of Physics, Tokyo Institute of Technology, 2-12-1
Ookayama, Meguro-ku, Tokyo 152-8551, Japan \and Physics
Department, Columbia University, 538 West 120th Street, New York,
NY 10027}
\abstract{We describe the weak localization correction to
conductivity in ultra-thin graphene films, taking into account
disorder scattering and the influence of trigonal warping of the
Fermi surface. A possible manifestation of the chiral nature of
electrons in the localization properties is hampered by trigonal
warping, resulting in a suppression of the weak anti-localization
effect in monolayer graphene and of weak localization in bilayer
graphene. Intervalley scattering due to atomically sharp
scatterers in a realistic graphene sheet or by edges in a narrow
wire tends to restore weak localization resulting in negative
magnetoresistance in both materials.}
%
\maketitle
\section{Introduction}\label{secintro}
The chiral nature of quasiparticles in ultra-thin graphitic films
\cite{DiVincenzo,Semenoff,haldane88,AndoNoBS,McCann} recently
revealed in Shubnikov de Haas and quantum Hall effect measurements
\cite{NovScience,Novoselov,Zhang,QHEbi} originates from the
hexagonal lattice structure of a monolayer of graphite (graphene).
The low energy behavior of monolayer graphene is explained in
terms of two valleys of Dirac-like chiral quasiparticles with
`isospin' linked to the momentum direction, exhibiting Berry phase
$\pi$ \cite{DiVincenzo,Semenoff,haldane88,AndoNoBS}. Remarkably,
the dominant low energy quasiparticles in a bilayer are different:
massive chiral quasiparticles with a parabolic dispersion and
Berry phase $2\pi$ \cite{McCann}.

In existing graphene structures, scattering occurs predominantly
from potential perturbations which are smooth on the scale of the
lattice constant $a$. This smooth potential arises from charges
located in the substrate at a distance $d$ from the 2D sheet,
$a\ll d<h/p_{F}$ ($h/p_{\mathrm{F}}$ being the Fermi wavelength).
Such a smooth potential is unable to change the isospin of chiral
electrons so that, in a monolayer, there is a complete suppression
of electron backscattering from potential disorder
\cite{AndoNoBS,AndoWL}. In the theory of quantum transport in
disordered systems \cite{WL} the suppression of backscattering is
known as the anti-localization (WAL) effect \cite{WLso} and, in
monolayer graphene with purely potential scattering, a possible
WAL behavior of conductivity \cite{AndoWL,khve06,guinea06,WLmono}
has been related to the Berry phase $\pi$ specific to the
Dirac-like Hamiltonian. Owing to the different degree of chirality
in bilayer graphene, related to Berry phase $2\pi$ \cite{McCann},
purely potential scattering would have a different effect: no
suppression of backscattering leading to conventional weak
localization (WL) \cite{guinea06,WLbilayer}.

In realistic graphene, there are other considerations that appear,
at first glance, to be merely small perturbations to this picture,
but they have a profound impact on the localization properties if
their effect is perceptible on length scales less than the phase
coherence length. This includes influence of ripples on the
graphene sheet \cite{WLman} leading to a weak randomization of
carbon $\pi$-bands, scattering off short-range defects that do not
conserve isospin and valley \cite{AndoWL,khve06,guinea06,WLmono},
and trigonal warping of the electronic band structure which
introduces asymmetry in the shape of the Fermi surface about each
valley \cite{AndoNoBS,WLmono,WLbilayer}. Both of them tend to
destroy the manifestation of chirality in the localization
properties, resulting in a suppression of the WAL effect in
monolayer graphene \cite{WLman} and of WL in bilayers
\cite{exeter}. Moreover, owing to the inverted chirality of
quasiparticles in the two valleys, intervalley scattering will
wash out any Berry phase effect and restore conventional weak
localization (WL) behavior of electrons in both monolayers and
bilayers in the regime of long-lasting phase coherence
\cite{AndoWL,khve06,guinea06,WLmono,WLbilayer,WLman,exeter,berger}.

\begin{figure}
\begin{center}\resizebox{0.7\columnwidth}{!}{\includegraphics{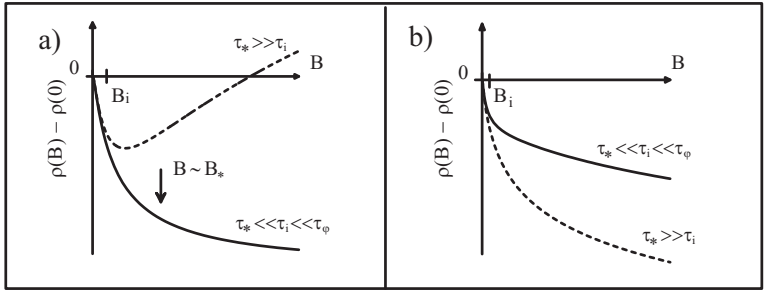}}
\caption{{($\mathbf{a}$) Typical magnetoresistance behavior
expected in a phase-coherent ($\tau_{\varphi} \gg \tau_i$)
monolayer of graphene for a weak intervalley scattering,
$\protect\tau_{\ast }\ll \protect\tau _{i}$ (solid line) and for
the case when the symmetry-breaking intravalley scattering is
slower than the intervalley one $\protect\tau _{\ast }\gg
\protect\tau _{i}$ (dashed). In both cases, we assume that the
phase coherence time determines the longest relaxation time scale
in the system. ($\mathbf{b}$) Magnetoresistance of bilayer
graphene, $\protect\tau_{\ast }\ll \protect\tau _{i}$ (solid line)
and $\protect\tau _{\ast }\gg \protect\tau _{i}$ (dashed). Note
that, for the case when $\tau_{\varphi} \ll \tau_i$ and
$\tau_{\ast} \ll \tau_i, \Delta \rho (B) =0$ in both monolayer and
bilayer graphene.}} \label{fig:1}
\end{center}
\end{figure}

Two typical magnetoresistance curves for monolayer graphene are
sketched in figure~\ref{fig:1}(a). They illustrate two extremes:
$\tau _{\ast } \gg \tau_{\mathrm{i}}$ and $\tau _{\ast } \ll
\tau_{\mathrm{i}}$ where $\tau_{\mathrm{i}}$ is the intervalley
scattering time and $\tau_{\ast}$ is the combined scattering time of
intravalley and intervalley scattering and of trigonal warping (see
equation~(\ref{tau_ast_mono},\ref{tau_int_mono}) and (\ref{lowgap})
below). When $\tau _{\ast } \gg \tau_{\mathrm{i}}$, the
magnetoresistance $\rho(B)-\rho(0)$ changes sign at the field
$B_{i}$ such that $\tau_{B}\sim \tau_{\mathrm{i}}$: from negative at
$B<B_{i}$ to positive at higher fields. This behavior resembles the
low-to-high field crossover in the quantum correction to the
conductivity of metals with strong spin-orbit coupling, though with
an inverted sign of the effect. In the case of $\tau_{\ast } \ll
\tau_{\mathrm{i}}$, the magnetoresistance is typically of a WL type,
with almost no sign of anti-localization up to the highest fields,
which shows that, unlike in a ballistic regime or a quantizing
magnetic field \cite{haldane88,McCann,CheianovFalko}, the chiral
nature of quasiparticles does not manifest itself in the weak field
magnetoresistance of realistic graphene structures. In bilayer,
however, slight enhancement of WL behavior is expected in the case
of weak intravalley symmetry breaking scattering, $\tau _{\ast} \ll
\tau_{\mathrm{i}}$, due to different Berry phase $2\pi$. In the case
of very strong intravalley symmetry breaking scattering, $\tau
_{\ast } \gg \tau_{\mathrm{i}}$ conventional WL magnetoresistance is
expected.

The WL behavior in graphene is novel because, with the exception
of spin-orbit coupling \cite{WLso,dresselhaus92}, qualitative
features of WL do not usually depend on the detail of the
electronic band structure and crystalline symmetry. In gapful
multi-valley semiconductors only the size of WL effect may depend
on the number of valleys and the strength of intervalley
scattering \cite{fuku80,bishop80,prasad95}. The low-field MR,
$\Delta \rho (B)\equiv \rho (B)-\rho (0)$, in a two dimensional
electron gas or a thin metallic film
\cite{WL,WLso,fuku80,socomment} in the absence of spin-orbit
coupling is characterized by
\begin{eqnarray}
\Delta \rho (B) &=& -\frac{s_{\theta} e^{2}\rho ^{2}}{2\pi
h}F\left( \frac{B}{B_{\varphi }} \right) ,\;B_{\varphi
}=\frac{\hbar c}{4De}\tau _{\varphi }^{-1} .  \label{summary}
\end{eqnarray}
Here $F(z)=\ln z+\psi (\frac{1}{2}+\frac{1}{z})$, $\tau_{\varphi}$
is the coherence time, $D$ is the diffusion coefficient, and the
integer factor $s_{\theta}$ depends on whether or not states in
$n_v$ valleys are mixed by disorder. This factor is controlled by
the ratio $\theta = \tau _{\mathrm{i}} / \tau_{\varphi}$ between
the intervalley scattering time $\tau_{\mathrm{i}}$ and the
coherence time $\tau_{\varphi}$. In materials such as Mg, ZnO, Si,
Ge, listed in Table~I, where each of the Fermi surface pockets is
$\mathbf{p}\rightarrow -\mathbf{p}$\ symmetric, intervalley
scattering reduces the size of the WL MR from that described by
$s_{\infty} = 2n_{v}$ when $\theta = \tau_{\mathrm{i}} /
\tau_{\varphi} \gg 1$ to $s_0 = 2$ for $\theta \ll 1$.

A more interesting scenario develops in a multi-valley semimetal,
where the localization properties can be influenced by the absence
of $\mathbf{p} \rightarrow - \mathbf{p}$\ symmetry of the electronic
dispersion within a single valley, and graphene is an example of
such a system. Here, we demonstrate how the asymmetry in the shape
of the Fermi surface in each of its two valleys determines the
observable WL behavior sketched in figure~\ref{fig:1}(a,b). It has a
tendency opposite to that known in usual semiconductors and metals:
a complete absence of WL MR for infinite $\tau_{\mathrm{i}}$
($s_{\infty} = 0$) and the standard WL effect in the limit of
$\tau_{\mathrm{i}} \, \ll \tau_{\varphi}$ ($s_0 = 2$). In
Section~\ref{secMRmono} we describe the WL effect in monolayer
graphene with a description of the low energy Hamiltonian in
Section~\ref{subsec:h1}, a qualitative account of interference
effects in Section~\ref{subsec:qual1}, the model of disorder in
Section~\ref{subsec:valley1}, an account of our diagrammatic
calculation of the weak localization correction in
Section~\ref{subsec:diagram1} and the resulting magnetoresistance in
Section~\ref{subsec:magneto1}. Section~\ref{secMRbi} describes the
weak localization correction and magnetoresistance in bilayer
graphene.

\begin{table}[tbp]
\caption{Weak localization factor $s_{\theta}$ in conductors with
a multi-valley conduction band and negligible spin-orbit coupling.
The factor $s_{\theta}$ is specified for two limiting cases, no
inter-valley scattering $\theta = \tau_{\mathrm{i}} /
\tau_{\varphi} \rightarrow \infty$, and for fast inter-valley
scattering $\theta \rightarrow 0$.}
\begin{center}
\begin{tabular}{clccc}
\hline \hspace{0.2cm}$n_{v}$\hspace{0.2cm} &
& \hspace{0.2cm}$s_{\infty }$\hspace{0.2cm} & \hspace{0.2cm}$s_{0}$\hspace{0.2cm}\\
\hline $1$ & \multicolumn{1}{l}{Mg films \cite{bergmann84}, ZnO
wells \cite{goldenblum99}}
& $2$ & - \\
$2,6$ & \multicolumn{1}{l}{Si MOSFETs \cite{fuku80,bishop80}} &
$2n_{v}$ & $2$ \\
$2$ & \multicolumn{1}{l}{Si/SiGe wells \cite{prasad95}} & $4$ & $2$ \\
$2$ & \multicolumn{1}{l}{monolayer graphene} & $0$ & $2$ \\
$2$ & \multicolumn{1}{l}{bilayer graphene} & $0$ & $2$ \\
 \hline
\end{tabular}
\end{center}
\end{table}

\section{Weak localization magnetoresistance in disordered monolayer graphene}\label{secMRmono}
\subsection{Low energy Hamiltonian of clean monolayer
graphene}\label{subsec:h1}
\begin{figure}
\begin{center}\resizebox{0.7\columnwidth}{!}{\includegraphics{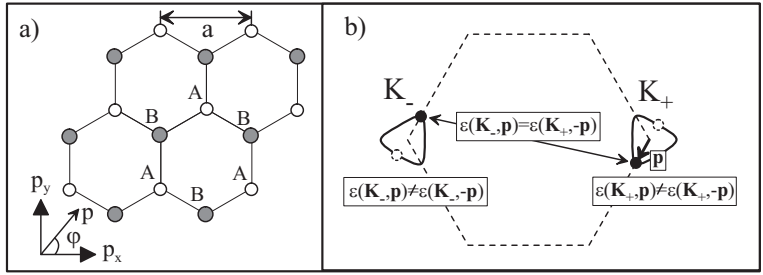}
} \caption{($\mathbf{a}$)Schematic plan view of the monolayer
lattice containing two sites in the unit cell, $A$ (white circles)
and $B$ (grey), arranged on an hexagonal lattice (solid lines).
($\mathbf{b}$)Fermi lines (solid lines) in the vicinity of two
inequivalent valleys $\mathbf{K}_{+}$ and $\mathbf{K}_{-}$ of the
hexagonal Brillouin zone (dashed line). Trigonal warping produces
asymmetry of the dispersion at each valley $\protect\epsilon
(\mathbf{K}_{\pm },\mathbf{p})\neq \protect \epsilon
(\mathbf{K}_{\pm },-\mathbf{p})$, where momentum $\mathbf{p}$ is
determined with respect to the center of the valley, but the
effects of warping in the valleys have opposite signs,
$\protect\epsilon (\mathbf{K}_{\pm },\mathbf{p} )=\protect\epsilon
(\mathbf{K}_{\mp },-\mathbf{p})$.} \label{fig:warping}
\label{fig:lattice1}
\end{center}
\end{figure}

The hexagonal lattice of monolayer graphene contains two
non-equivalent sites $A$ and $B$ in the unit cell, as shown in
figure~\ref{fig:lattice1}(a). The Fermi level in a neutral graphene
sheet is pinned near the corners of the hexagonal Brillouin zone
with wave vectors $\mathbf{K}_{\pm } = \pm
({\textstyle\frac{2}{3}}ha^{-1},0)$ where $a$ is the lattice
constant. The Brillouin zone corners $\mathbf{K}_{\pm }$ determine
two non-equivalent valleys in the quasiparticle spectrum described
by the Hamiltonian \cite{DiVincenzo,AndoNoBS,WLmono,wallace},
\begin{eqnarray}
{\hat{H}}_{1} &=& v \Pi _{z} \left( \sigma_{x} p_{x}+\sigma_{y}
p_{y}\right)
+ {\hat{h}}_{1w},  \label{h1} \\
{\hat{h}}_{1w} &=& \mu \Pi_{0} \left[ \sigma_{y} \left(
{p}_{x}{p}_{y}+{p}_{y}{p}_{x}\right) - \sigma_{x} \left(
{p}_{x}^{2}-{p}_{y}^{2}\right) \right] .  \nonumber
\end{eqnarray}
This Hamiltonian operates in the space of four-component wave
functions, $\Phi = [\phi _{\mathbf{K}_{+}}(A),
 \phi_{\mathbf{K}_{+}}(B), \\ \phi_{\mathbf{K}_{-}}(B),
\phi_{\mathbf{K}_{-}}(A)] $ describing electronic amplitudes on
$A$ and $B$ sites and in the valleys $\mathbf{K}_{\pm }$. Here, we
use a direct product of `isospin' ($AB$ lattice space) matrices
$\sigma _{0}\equiv \hat{1},\sigma _{x,y,z}$ and `pseudospin'
inter/intra-valley matrices $\Pi _{0}\equiv \hat{1},\Pi _{x,y,z}$
to highlight the difference between the form of ${\hat{H}}_{1}$ in
the non-equivalent valleys. The Hamiltonian ${\hat{H}}_{1}$ takes
into account nearest neighbor $A/B$ hopping in the plane with the
first (second) term representing the first (second) order term in
an expansion with respect to momentum $\mathbf{p}$ measured from
the center of the valley $\mathbf{K}_{\pm }$.

Near the center of the valley $\mathbf{K}_{+}$, the Dirac-type
part, $v\,\mathbf{\sigma \cdot p}$, of ${\hat{H}}_{1}$ determines
the linear dispersion $\epsilon=vp$ for the electron in the
conduction band and $\epsilon=-vp$ for the valence band. Electrons
in the conduction and valence band also differ by the isospin
projection onto the direction of their momentum (chirality):
$\mathbf{\sigma }\cdot \mathbf{p}/p=1$ in the conduction band,
$\mathbf{\sigma }\cdot \mathbf{p}/p=-1$ in the valence band. In
the valley $\mathbf{K}_{-}$, the electron chirality is
mirror-reflected: it fixes $\mathbf{\sigma }\cdot \mathbf{p}/p =
-1$ for the conduction band and $\mathbf{\sigma }\cdot
\mathbf{p}/p = 1$ for the valence band. For an electron in the
conduction band, the plane wave state is
\begin{eqnarray}
\Phi_{\mathbf{K}_\pm,\mathbf{p}} &=& \frac{e^{ i\mathbf{p
r}/\hbar}}{\sqrt{2}} \left(\pm e^{-i\varphi /2}|\uparrow
\rangle_{\mathbf{K}_\pm,\mathbf{p}} + e^{i\varphi /2}|\downarrow
\rangle_{\mathbf{K}_\pm,\mathbf{p}} \right) , \label{pl1} \\
\Phi_{\mathbf{K}_\pm,-\mathbf{p}} &=& \frac{i e^{- i\mathbf{p
r}/\hbar}}{\sqrt{2}} \left( \mp e^{-i\varphi /2}|\uparrow
\rangle_{\mathbf{K}_\pm,-\mathbf{p}}+ e^{i\varphi /2}|\downarrow
\rangle_{\mathbf{K}_\pm,-\mathbf{p}} \right). \label{pl2}
\end{eqnarray}
Here ${|\uparrow \rangle_{\mathbf{K}_{+},\mathbf{p}}}=[1,0,0,0]$,
${|\downarrow \rangle_{\mathbf{K}_{+},\mathbf{p}}}=[0,1,0,0]$ and
${|\uparrow \rangle_{\mathbf{K}_{-},\mathbf{p}}}=[0,0,1,0]$,
${|\uparrow \rangle_{\mathbf{K}_{-},\mathbf{p}}}=[0,0,0,1]$, and the
factors $e^{\pm i\varphi /2}$ take into account the chirality, with
angle $\varphi$ defining the direction of momentum in the plane
$\mathbf{p}=(p\cos\varphi ,p\sin\varphi)$. The angular dependence $w
( \varphi ) \sim \cos^2 ( \varphi /2 )$ of the scattering
probability off a short range potential which conserves isospin is
shown in figure~\ref{fig:monopath}(a). It demonstrates the fact that
the chiral states Eqs.~(\ref{pl1},\ref{pl2}) with isospin fixed to
the direction of momentum display an absence of back scattering
\cite{AndoNoBS,shon98,AndoWL}, leading to a transport time longer
than the scattering time $\tau_{\textrm{tr}} = 2 \tau_0$.

The term ${\hat{h}}_{1\mathrm{w}}$ in equation~(\ref{h1}) can be
treated as a perturbation leading to a trigonal deformation of a
single-connected Fermi line and $\mathbf{p}\rightarrow -\mathbf{p}$\
asymmetry of the electron dispersion inside each valley illustrated
in figure~\ref{fig:warping}(b): $\epsilon (\mathbf{K}_{\pm
},\mathbf{p})\neq \epsilon (\mathbf{K}_{\pm}, -\mathbf{p})$.
However, due to time-reversal symmetry \cite{t-rev} trigonal warping
has opposite signs in the two valleys and $\epsilon (\mathbf{K}_{\pm
},\mathbf{p})= \epsilon (\mathbf{K}_{\mp },-\mathbf{p})$. The
interplay between the two terms in ${\hat{H}}_{1}$ resulting in the
asymmetry of the electronic dispersion manifest itself in the WL
behavior.

\subsection{Interference of electronic waves in monolayer graphene}\label{subsec:qual1}
\begin{figure}
\begin{center}\resizebox{0.6\columnwidth}{!}{\includegraphics{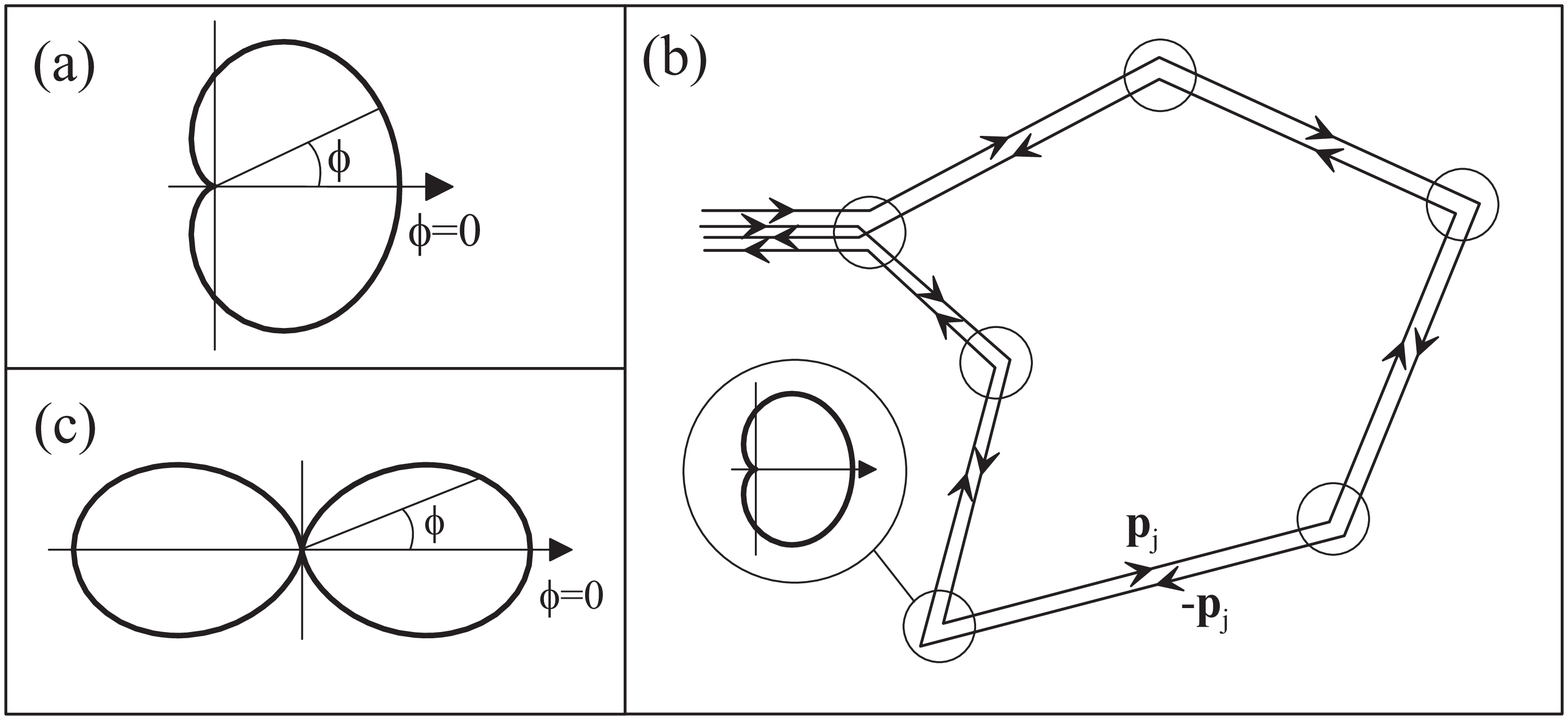}}
\caption{(a) angular dependence $w ( \varphi ) \sim \cos^2 (
\varphi /2 )$ of the scattering probability off a short range
potential in monolayer graphene, (b) a pair of closed paths which
contribute to weak localization, (c) angular dependence $w (
\varphi ) \sim \cos^2 ( \varphi )$ of the scattering probability
off a short range potential in bilayer graphene.}
\label{fig:monopath}
\end{center}
\end{figure}

The WL correction to conductivity in disordered conductors is a
result of the constructive interference of electrons propagating
around closed loops in opposite directions \cite{WL} as sketched in
figure~\ref{fig:monopath}(b). Such interference is constructive in
metals and semiconductors with negligibly weak spin-orbit coupling,
since electrons acquire exactly the same phase when travelling along
two time-reversed trajectories.

WL is usually described \cite{WL} in terms of the particle-particle
correlation function, Cooperon. Following the example of Cooperons
for a spin ${\textstyle \frac{1}{2}}$, we classify Cooperons as
singlets and triplets in terms of `isospin' ($AB$ lattice space) and
`pseudospin' (inter/intra-valley) indices (see
Section~\ref{subsec:h1}). In fact, with regards to the isospin
(sublattice) composition of Cooperons in a disordered monolayer,
only singlet modes are relevant. This is because a correlator
describing two plane waves, $\Phi _{\mathbf{K}_+,\mathbf{p}}$ and
$\Phi_{\mathbf{K}_-,-\mathbf{p}}$ Eqs.~(\ref{pl1},\ref{pl2}),
propagating in opposite directions along a ballistic segment of a
closed trajectory as in figure~\ref{fig:monopath}(b) has the
following form:
\begin{eqnarray*}
\Phi _{\mathbf{K},\mathbf{p}}\Phi
_{\mathbf{K}^{\prime},-\mathbf{p}} &\sim & |\uparrow
\rangle_{\mathbf{K},\mathbf{p}} |\downarrow
\rangle_{\mathbf{K}^{\prime},-\mathbf{p}} - |\downarrow
\rangle_{\mathbf{K},\mathbf{p}} |\uparrow
\rangle_{\mathbf{K}^{\prime},-\mathbf{p}} - e^{-i\varphi}
|\uparrow \rangle_{\mathbf{K},\mathbf{p}} |\uparrow
\rangle_{\mathbf{K}^{\prime},-\mathbf{p}} + e^{i\varphi}
|\downarrow \rangle_{\mathbf{K},\mathbf{p}} |\downarrow
\rangle_{\mathbf{K}^{\prime},-\mathbf{p}} .
\end{eqnarray*}
It contains only sublattice-singlet terms (the first two terms)
because triplet terms (the last two terms) disappear after
averaging over the direction of momentum, $\mathbf{p}=(p\cos
\varphi, p\sin\varphi)$, so that $\langle e^{\pm i\varphi }\rangle
_{\varphi }=0$. In fact, our diagrammatic calculation described in
Section~\ref{subsec:diagram1} shows that the interference
correction to the conductivity of graphene is determined by the
interplay of four isospin singlet modes: one pseudospin singlet
and three pseudospin triplets. Of these, two of the pseudospin
triplet modes are intravalley Cooperons while the remaining
triplet and the singlet are intervalley Cooperons.

In the WL picture for a diffusive electron in a metal, two phases
$\vartheta _{1}$ and $\vartheta _{2}$ acquired while propagating
along paths $"1"$ and $"2"$ [see figure~\ref{fig:monopath}(b)] are
exactly equal, so that the interference of such paths is
constructive and, as a result, enhances backscattering leading to WL
\cite{WL}. In monolayer graphene the Berry phase $\pi$
characteristic for quasi-particles described by the first term of
${\hat{H}}_{1}$, determines the phase difference $\delta \equiv
\vartheta_{1} - \vartheta_{2} = \pi N$ (where $N$ is the winding
number of a trajectory) \cite{AndoWL,WLmono}, and one would expect
weak anti-localization behavior. However, the asymmetry of the
electron dispersion due to ${\hat{h}}_{\mathrm{1w}}$, leading to
warping of the Fermi line around each valley as in
figure~\ref{fig:warping}(b), deviates $\delta$ from $\pi N$. Indeed,
any closed trajectory is a combination of ballistic intervals,
figure~\ref{fig:monopath}(b). Each interval, characterized by the
momenta $\pm \mathbf{p_j}$ (for the two directions) and by its
duration $t_j$, contributes to the phase difference $\delta_j =
[\epsilon(\mathbf{p_j}) - \epsilon(-\mathbf{p_j})]t_j =
{\hat{h}}_{\mathrm{1w}}(\mathbf{p_j})t_j$. Since $\delta_j$ are
random uncorrelated, the mean square of $\delta = \sum\delta_j$ can
be estimated as $\langle\delta^2\rangle \sim
\langle(t_j{\hat{h}}_{\mathrm{1w}}(\mathbf{p_j}))^2\rangle
t/\tau_{tr}$, where $t$ is the duration of the path and $\tau_{tr}$
is the transport mean free time. Warping thus determines the
relaxation rate,
\begin{equation}
\tau _{\mathrm{w}}^{-1} \sim \langle
\mathrm{Tr}{\hat{h}}_{\mathrm{1w}}^{2}(\mathbf{p})\rangle_{\varphi},
\label{tauw1}
\end{equation}
which suppresses the two intravalley Cooperons, and, thus, weak
anti-localization in the case when electrons seldom change their
valley state. The two intervalley Cooperons are not affected by
trigonal warping due to time-reversal symmetry of the system which
requires $\epsilon (\mathbf{K}_{\pm },\mathbf{p})=\epsilon
(\mathbf{K}_{\mp },-\mathbf{p})$, figure~\ref{fig:warping}(b). These
two Cooperons cancel each other in the case of weak intervalley
scattering, thus giving $\delta g \sim 0$. However, intervalley
scattering, with a rate $\tau _{\mathrm{i}}^{-1}$ larger than the
decoherence rate $\tau _{\varphi }^{-1}$, breaks the exact
cancellation of the two intervalley Cooperons and partially restores
weak localization.

\subsection{Matrix parameterization, valley symmetry and the model of
disorder}\label{subsec:valley1}
To describe the valley symmetry of monolayer graphene and
parameterize all possible types of disorder, we introduce two sets
of 4$\times $4 Hermitian matrices $\vec{\Sigma} =(\Sigma
_{x},\Sigma _{y},\Sigma _{z})$ with $[\Sigma _{s_{1}},\Sigma
_{s_{2}}]=2i\varepsilon ^{s_{1}s_{2}s_{3}}\Sigma _{s_{3}}$, and
'pseudospin' $\vec{\Lambda}=(\Lambda _{x},\Lambda _{y},\Lambda
_{z})$ with $[\Lambda _{l_{1}},\Lambda _{l_{2}}]=2i\varepsilon
^{l_{1}l_{2}l_{3}}\Sigma _{l_{3}}$,
defined as%
\begin{eqnarray}
\Sigma _{x} &=&\Pi _{z}\otimes \sigma _{x},\;\Sigma _{y}=\Pi
_{z}\otimes
\sigma _{y},\;\Sigma _{z}=\Pi _{0}\otimes \sigma _{z},  \label{Sigma} \\
\Lambda _{x} &=&\Pi _{x}\otimes \sigma _{z},\;\Lambda _{y}=\Pi
_{y}\otimes \sigma _{z},\;\Lambda _{z}=\Pi _{z}\otimes \sigma
_{0}.  \label{Lamda}
\end{eqnarray}%
The operators $\vec{\Sigma}$ and $\vec{\Lambda}$ form two mutually
independent algebras equivalent to the algebra of Pauli matrices (in
Eqs. (\ref{Sigma},\ref{Lamda}) $\varepsilon ^{s_{1}s_{2}s_{3}}$ is
the antisymmetric tensor and $[\Sigma_{s} ,\Lambda _{l}]=0$) thus
they determine two commuting subgroups of the group U$_{4}$ of
unitary transformations \cite{U4} of a 4-component $\Phi $: an
'isospin' sublattice group SU$_{2}^{\Sigma }\equiv
\{e^{ia\vec{n}\cdot \!\vec{\Sigma}}\}$ and a 'pseudospin' valley
group SU$_{2}^{\Lambda }\equiv \{\mathrm{e}^{ib\vec{n} \cdot
\!\vec{\Lambda}}\}$. Also, $\vec{\Sigma}$ and $\vec{\Lambda}$ change
sign under the inversion of time, whereas products $\Sigma
_{s}\Lambda _{l}$ are invariant with respect to the $t\rightarrow
-t$ transformation and can be used as a basis for a quantitative
phenomenological description of non-magnetic static disorder
\cite{shon98,mf05}. Table~1 is a summary of the discrete symmetries
of the operators $\vec{\Sigma}$ and $\vec{\Lambda}$ and their
products $\Sigma_{s}\Lambda_{l}$. Time reversal $T$ of an operator
$\hat{W}$ is described by $(\Pi _{x}\otimes
\sigma_{x}){\hat{W}}^{\ast }(\Pi _{x}\otimes \sigma _{x})$. The
rotation by $\pi /3$ about the perpendicular $z$ axis is described
by $C_6 = \Pi_{x} \otimes \exp [ (-2\pi i/3) \sigma_z ]$. Reflection
in the $x$-$z$ plane is $R_x = \Pi_{0} \otimes \sigma_x$.

The operators $\vec{\Sigma}$ and $\vec{\Lambda}$ help us to
represent the electron Hamiltonian in weakly disordered graphene
as
\begin{eqnarray}
{\hat{H}_1}=v\,\vec{\Sigma}\mathbf{p}+{\hat{h}}_{\mathrm{1w}}+
\mathrm{\hat{I}}u(\mathbf{r}) + \sum_{s,l=x,y,x}\Sigma_{s}\Lambda
_{l}u_{s,l}(\mathbf{r}),
\label{h1-2} \\
\mathrm{where}\;\;{\hat{h}}_{\mathrm{1w}}=- \mu \Sigma_{x}
(\,\vec{\Sigma} \mathbf{p}) \Lambda_{z} \Sigma_{x}
(\,\vec{\Sigma}\mathbf{p}) \Sigma_{x}. \nonumber
\end{eqnarray}%
The Dirac-type part $v\,\vec{\Sigma}\mathbf{p}$ of ${\hat{H}_1}$ in
equation~(\ref{h1-2}) and potential disorder
$\mathrm{\hat{I}}u(\mathbf{r})$ (where $\mathrm{\hat{I}}$ is a
4$\times $4 unit matrix and $\left\langle u\left( \mathbf{r}\right)
u\left( \mathbf{r}^{\prime }\right) \right\rangle =u^{2}\delta
\left( \mathbf{r}-\mathbf{r}^{\prime }\right) $) do not contain
valley operators $\mathbf{\Lambda }_{l}$, thus, they remain
invariant with respect to the pseudospin transformations from valley
group SU$_{2}^{\Lambda }$. Below, we assume that the
isospin/pseudospin-conserving disorder due to charges lying in a
substrate at distances from the graphene sheet shorter or comparable
to the electron wavelength $h/p_{\mathrm{F}}$ dominates the elastic
scattering rate, $\tau ^{-1}\approx \tau _{0}^{-1}=\pi \gamma
u^{2}/\hbar $, where $\gamma =p_{\mathrm{F}}/(2\pi \hbar ^{2}v)$ is
the density of states of quasiparticles per spin in one valley. All
other types of disorder which originate from atomically sharp
defects \cite{shon98,mf05} and break the SU$_{2}^{\Lambda }$
pseudospin symmetry of the system are included in a random matrix
$\Sigma _{s}\Lambda _{l}u_{s,l}(\mathbf{r})$. In particular,
$u_{z,z}(\mathbf{r})$ describes disorder due to
different on-site energies on the $A$ and $B$ sublattices, $u_{x(y),z}(%
\mathbf{r})$ plays the role of a valley-antisymmetric vector
potential of a geometrical nature, and $u_{s,x(y)}(\mathbf{r})$
take into account inter-valley scattering. For simplicity, we
assume that different types of disorder are uncorrelated, $\langle
u_{s,l}(\mathbf{r})u_{s^{\prime },l^{\prime }}(\mathbf{r}^{\prime
})\rangle =u_{sl}^{2}\delta _{ss^{\prime }}\delta _{ll^{\prime
}}\delta (\mathbf{r}-\mathbf{r}^{\prime })$ and, on average,
isotropic in the $x-y$ plane, $u_{xl}^{2}=u_{yl}^{2}\equiv u_{\bot
l}^{2}$, $u_{sx}^{2}=u_{sy}^{2}\equiv u_{s\bot }^{2}$. We
parametrize them by scattering rates $\tau _{sl}^{-1}=\pi \gamma
u_{sl}^{2}/\hbar $. Also, the warping term,
${\hat{h}}_{\mathrm{1w}}$ lifts the pseudospin symmetry
SU$_{2}^{\Lambda }$, though it remains invariant under pseudospin
rotations around the z-axis.

To characterize Cooperons in monolayer graphene, we use a Cooperon
matrix $C_{\alpha \beta \alpha ^{\prime }\beta ^{\prime }}^{\xi
\mu \xi ^{\prime }\mu ^{\prime }}$ where subscripts describe the
isospin state of incoming $\alpha \beta $ and outgoing $\alpha
^{\prime }\beta ^{\prime }$ pairs of electrons and superscripts
describe the pseudospin state of incoming $\xi \mu $ and outgoing
$\xi ^{\prime }\mu ^{\prime }$ pairs. Following the example of
Cooperons for a spin ${\textstyle \frac{1}{2}}$, we classify
Cooperons as singlets and triplets in terms of isospin and
pseudospin indices $C_{S_{1}S_{2}}^{M_{1}M_{2}}$. For example,
$M=0$ is a `pseudospin-singlet', $M=x,y,z$ are three
`pseudospin-triplet' components; $S=0$ is a `isospin-singlet' and
$S=x,y,z$ are `pseudospin-triplet' components. It is convenient to
use pseudospin as a quantum number to classify the Cooperons in
graphene because of the hidden SU$_{2}^{\Lambda }$ symmetry of the
dominant part of the free-electron and disorder Hamiltonian.

\begin{table}[tbp]
\caption{Matrices ${\Sigma}_s$ and ${\Lambda}_l$ provide us with
representations of the crystalline symmetry group, which is
constructed of $3$ generators, ${\textstyle\frac{\pi}{3}}$-rotation,
$C_6$, mirror reflection with respect to $Ox$ axis, $R_x$, and
translation along $Ox$ by lattice constant, $\mathbf{a}$. Operation
$T$ stands for the time reversal, $t \rightarrow -t$. Here,
${\Sigma}_s$ and ${\Lambda}_l$ are grouped into bases forming
irreducible representations which can be $1, 2$ and 4 dimensional.
The transformation matrixes $U_{ji}$, $g \phi_i = \sum_{j}
U_{ji}(g)\phi_j$ ($g$ stands for a symmetry operation), are given
explicitly for each of such bases $\phi_i$.}
\begin{center}
\begin{tabular}{|c|c|c|c|c|}
\hline\hline  $\Sigma _{s}\Lambda _{l}$ & $T$ &
  $C_{6}$ & $R_{x}$ &  $\mathbf{a}$  \\
\hline $\hat{I}$ & $+1$ & $+1$ & $+1$ & $+1$  \\
\hline
$\Sigma _{z}$ & $-1$ & $+1$ & $-1$ & $+1
$  \\
$\Lambda _{z}$ & $-1$ & $-1$ & $+1$ & $+1$ \\
$\Sigma _{z}\Lambda _{z}$ & $+1$ & $-1$ &
$-1$ & $+1$  \\
$\left[\begin{array}{cc} \Sigma_x & \\ \Sigma_y & \\
\end{array}\right]$ &  $-1$ &
$\left(\begin{array}{cc}
  \frac{1}{2} & \frac{\sqrt{3}}{2} \\
  -\frac{\sqrt{3}}{2} & \frac{1}{2} \\
  \end{array}\right)$ & $\left(
\begin{array}{cc}
  1 & 0 \\
  0 & -1 \\
  \end{array}
  \right)$ &  $\left(
\begin{array}{cc}
  1 & 0 \\
  0 & 1 \\
  \end{array}
  \right)$  \\
$ \left[
\begin{array}{cc}
  \Lambda_x  \\
  \Lambda_y   \\
  \end{array}
\right]$ & $-1$ & $\left(
\begin{array}{cc}
  1 & 0 \\
  0 & -1 \\
  \end{array}
  \right)$ & $\left(
\begin{array}{cc}
  -1 & 0 \\
  0 & -1 \\
  \end{array}
  \right)$ & $\left(
\begin{array}{cc}
   -\frac{1}{2} & -\frac{\sqrt{3}}{2} \\
  \frac{\sqrt{3}}{2} & -\frac{1}{2} \\
  \end{array}
  \right)$  \\
$\left[
\begin{array}{cc}
  \Lambda_z\Sigma_x  \\
  \Lambda_z\Sigma_y   \\
  \end{array}
\right]$ & $+1$ & $\left(
\begin{array}{cc}
  -\frac{1}{2} & -\frac{\sqrt{3}}{2} \\
  \frac{\sqrt{3}}{2} & -\frac{1}{2} \\
  \end{array}
  \right)$ & $\left(
\begin{array}{cc}
  1 & 0 \\
  0 & -1 \\
  \end{array}
  \right)$ & $\left(
\begin{array}{cc}
  1 & 0 \\
  0 & 1 \\
  \end{array}
  \right)$  \\
$\left[
\begin{array}{cc}
  \Lambda_x\Sigma_z  \\
  \Lambda_y\Sigma_z   \\
  \end{array}
\right]$ & $+1$ & $\left(
\begin{array}{cc}
  1 & 0 \\
  0 & -1 \\
  \end{array}
  \right) $ & $\left(
\begin{array}{cc}
  1 & 0 \\
  0 & 1 \\
  \end{array}
  \right)$ & $\left(
\begin{array}{cc}
   -\frac{1}{2} & -\frac{\sqrt{3}}{2} \\
  \frac{\sqrt{3}}{2} & -\frac{1}{2} \\
  \end{array}
  \right)$  \\
$\left[
\begin{array}{cccc}
  \Lambda_x\Sigma_x  \\
  \Lambda_x\Sigma_y  \\
  \Lambda_y\Sigma_x   \\
  \Lambda_y\Sigma_y   \\
  \end{array}
\nonumber \right]$ & $+1$ & $\left(
\begin{array}{cccc}
  \frac{1}{2} & \frac{\sqrt{3}}{2} & 0 & 0 \\
  -\frac{\sqrt{3}}{2} & \frac{1}{2} & 0 & 0 \\
  0 & 0 & -\frac{1}{2} & -\frac{\sqrt{3}}{2} \\
  0 & 0 &  \frac{\sqrt{3}}{2} & -\frac{1}{2} \\
  \end{array}
  \right)$ & $\left(
\begin{array}{cccc}
  -1 & 0 & 0 & 0 \\
  0 & 1 & 0 & 0 \\
  0 & 0 & -1 & 0 \\
  0 & 0 & 0 & 1 \\
  \end{array}
  \right)$ & $\left(
\begin{array}{cccc}
  -\frac{1}{2} & 0 & -\frac{\sqrt{3}}{2} & 0 \\
  0 &  -\frac{1}{2} & 0 & -\frac{\sqrt{3}}{2} \\
  \frac{\sqrt{3}}{2} & 0 & -\frac{1}{2} & 0 \\
  0 & \frac{\sqrt{3}}{2} & 0 & -\frac{1}{2} \\
  \end{array}
  \right)$  \\
\hline\hline
\end{tabular}
\end{center}
\end{table}

%
\subsection{Diagrammatic calculation of the weak localization
correction in monolayer graphene}\label{subsec:diagram1}
To describe the quantum transport of 2D electrons in graphene we
evaluate the disorder-averaged one-particle Green's functions,
vertex corrections, calculate the Drude conductivity and transport
time, classify Cooperon modes and derive equations for those which
are gapless in the limit of purely potential disorder. In
Section~\ref{subsec:magneto1} we analyse `Hikami boxes'
\cite{WL,WLso} for the weak localization diagrams paying attention
to a peculiar form of the current operator for Dirac electrons and
evalute the interference correction to conductivity leading to the
WL magnetoresistance. In these calculations, we treat trigonal
warping ${\hat{h}}_{\mathrm{1w}}$ in the free-electron Hamiltonian
Eqs. (\ref{h1},\ref{h1-2}) perturbatively, assume that potential
disorder $\mathrm{\hat{I}}u(\mathbf{r})$ dominates in the elastic
scattering rate, $\tau ^{-1}\approx \tau _{0}^{-1}=\pi \gamma
u^{2}/\hbar $, and take into account all other types of disorder
when we determine the relaxation spectra of low-gap Cooperons.

\begin{figure}
\begin{center}\resizebox{0.60\columnwidth}{!}{\includegraphics{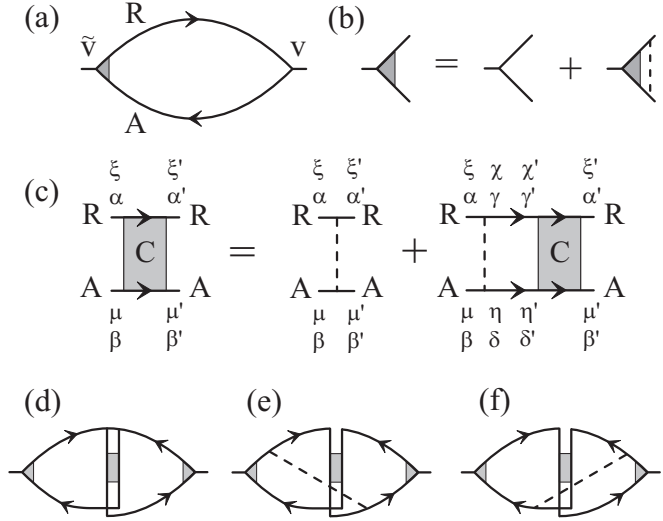}}
\caption{(a) Diagram for the Drude conductivity with (b) the
vertex correction. (c) Bethe-Salpeter equation for the Cooperon
propagator with
valley indices $\protect\xi \protect\mu \protect\xi ^{\prime }\protect\mu %
^{\prime }$ and AB lattice indices $\protect\alpha \protect\beta \protect%
\alpha ^{\prime }\protect\beta ^{\prime }$. (d) Bare 'Hikami box'
relating the conductivity correction to the Cooperon propagator
with (e) and (f) dressed 'Hikami boxes'. Solid lines represent
disorder averaged $G^{R/A}$, dashed lines represent disorder.}
\label{fig:2}
\end{center}
\end{figure}

Using the standard methods of the diagrammatic technique for
disordered systems \cite{WL,WLso} and assuming that
$p_{\mathrm{F}}v\tau \gg \hbar $, we obtain the disorder averaged
single particle Green's function,
\begin{equation}
\hat{G}^{R/A}\left( \mathbf{p},\epsilon \right) =\frac{\epsilon
_{R/A}+v\, \vec{\Sigma}\mathbf{p}}{\epsilon
_{R/A}^{2}-v^{2}p^{2}},\;\;\epsilon _{R/A}=\epsilon \pm
{\textstyle\frac{1}{2}}i\hbar \tau _{0}^{-1}.  \nonumber
\end{equation}
Note that, for the Dirac-type particles described in
equation~(\ref{h1}), the current operator is a momentum-independent
matrix vector, $\mathbf{\hat{v}}=v \vec{\Sigma}$. As a result, the
current vertex $\tilde{v}_{j}$ ( $j=x,y$), which appears as a block
in figure \ref{fig:2}(a) describing the Drude conductivity,
\begin{eqnarray}
g_{jj} = \frac{e^{2}}{\pi \hbar }\int \frac{d^{2}p}{\left( 2\pi
\right)^{2}} \mathrm{Tr}\left\{ \tilde{v}_{j}\hat{G}^{R} \left(
\mathbf{p},\epsilon \right) \hat{v}_{j}\hat{G}^{A}\left(
\mathbf{p},\epsilon \right) \right\} = 4e^{2}\gamma D ; \qquad
D=v^{2}\tau _{0}\equiv {\textstyle\frac{1}{2}}v^{2}\tau
_{\mathrm{tr}}, \label{conductivity}
\end{eqnarray}
is renormalised by vertex corrections \cite{shon98} in
figure~\ref{fig:2}(b): $\mathbf{\tilde{v}} = 2 \mathbf{\hat{v}}= 2 v
\vec{\Sigma}$. Here `$\mathrm{Tr}$' stands for the trace over the AB
and valley indices. Using the Einstein relation in equation
(\ref{conductivity}), we see that due to the anisotropy of
scattering [\textit{i.e.}, lack of backscattering from an individual
Coulomb centre as in figure~\ref{fig:monopath}(a)] the transport
time in graphene is twice the scattering time, $\tau
_{\mathrm{tr}}=2\tau _{0}$. Note that in
equation~(\ref{conductivity}) spin degeneracy has been taken into
account.


The Cooperon $C_{\alpha \beta ,\alpha ^{\prime }\beta ^{\prime
}}^{\xi \mu ,\xi ^{\prime }\mu ^{\prime }}$ obeys the Bethe-Salpeter
equation represented diagrammatically in figure~\ref{fig:2}(c). The
shaded blocks in figure~\ref{fig:2}(c) are infinite series of ladder
diagrams, while the dashed lines represent the correlator of the
disorder in equation (\ref{h1-2}). We classify Cooperons in graphene
as iso- and pseudospin singlets and triplets, as was mentioned
above, with the help of the following relation,
\begin{eqnarray}
C_{s_{1}s_{2}}^{l_{1}l_{2}} =\frac{1}{4} \sum_{\alpha ,\beta
,\alpha ^{\prime },\beta ^{\prime },} \sum_{\xi ,\mu ,\xi ^{\prime
},\mu ^{\prime },} \left( \Sigma _{y}\Sigma _{s_{1}} \Lambda
_{y}\Lambda _{l_{1}}\right)_{\alpha \beta }^{\xi \mu }C_{\alpha
\beta ,\alpha ^{\prime }\beta ^{\prime }}^{\xi \mu ,\xi ^{\prime
}\mu ^{\prime }} \left( \Sigma_{s_{2}}\Sigma_{y} \Lambda_{l_{2}}
\Lambda_{y} \right)_{\beta ^{\prime }\alpha ^{\prime }}^{\mu
^{\prime }\xi^{\prime }}\,. \label{Cspin}
\end{eqnarray}
Such a classification of modes is permitted by the commutation of
the iso- and pseudospin operators $\vec{\Sigma}$ and
$\vec{\Lambda}$ in Eqs. (\ref{Sigma},\ref{Lamda},\ref{Cspin}),
$[\Sigma _{s},\Lambda _{l}]=0$. To select the isospin singlet
($s=0$) and triplet ($s=x,y,z$) Cooperon components (scalar and
vector representation of the sublattice group SU$_{2}^{\Sigma
}\equiv \{e^{ia\vec{n}\cdot \!\vec{\Sigma}}\}$), we project the
incoming and outgoing Cooperon indices onto matrices $\Sigma
_{y}\Sigma _{s_{1}}$and $\Sigma _{s_{2}}\Sigma _{y}$,
respectively. The pseudospin singlet ($l=0$) and triplet
($l=x,y,z$) Cooperons (scalar and vector representation of the
valley group SU$_{2}^{\Lambda }\equiv \{\mathrm{e}^{ib\vec{n}\cdot
\!\vec{\Lambda}}\}$) are determined by the projection of
$C_{\alpha \beta ,\alpha ^{\prime }\beta ^{\prime }}^{\xi \mu ,\xi
^{\prime }\mu ^{\prime }}$ onto matrices $\Lambda _{y}\Lambda
_{l_{1}}$($\Lambda _{l_{2}}\Lambda _{y}$) and are accounted for by
superscript indices in $C_{s_{1}s_{2}}^{l_{1}l_{2}}$.

For 'diagonal' disorder $\mathrm{\hat{I}}u(\mathbf{r})$, the
Bethe-Salpeter equation, figure \ref{fig:2}(c) takes the form
\begin{eqnarray}
C_{s_{1}s_{2}}^{l_{1}l_{2}}\left( \mathbf{q}\right)&=&\tau
_{0}\,\delta ^{l_{1}l_{2}}\delta _{s_{1}s_{2}} \\ &+&
\frac{1}{4\pi \gamma \tau _{0}\hbar
}\sum_{s,l}C_{ss_{2}}^{ll_{2}}\left( \mathbf{q}\right) \int
\frac{d^{2}p}{\left( 2\pi \right) ^{2}} \mathrm{Tr}\Big\{\Sigma
_{s}\Sigma_{y}\Lambda_{l}\Lambda_{y} \left[
\hat{G}_{\mathbf{p},\hbar \omega +\epsilon }^{R}\right]
^{\mathrm{t}}\Lambda _{y}\Lambda _{l_{1}}\Sigma _{y}\Sigma
_{s_{1}}\hat{G}_{\hbar \mathbf{q}-\mathbf{p},\epsilon }^{A}\Big\}.
\nonumber
\end{eqnarray}
It leads to a series of coupled equations for the Cooperon modes
$C_{ss}^{ll}\equiv C_{s}^{l}$. It turns out that for potential
disorder~$\mathrm{\hat{I}}u(\mathbf{r})$ isospin-singlet modes
$C_{0}^{l}$ are gapless in all (singlet and triplet) pseudospin
channels, whereas triplet modes $C_{x}^{l}$ and $C_{y}^{l}$ have
relaxation gaps $\Gamma _{x}^{l}=\Gamma _{y}^{l}=\frac{1}{2}\tau
_{0}^{-1}$ and $C_{z}^{l}$ have gaps $\Gamma _{z}^{l}=\tau
_{0}^{-1}$. When obtaining the diffusion equations for the Cooperons
using the gradient expansion of the Bethe-Salpeter equation we take
into account its matrix structure. We find that {isospin-singlets
}$C_{0}^{l}${\ are coupled to the triplets }$C_{x}^{l}$ and
$C_{y}^{l}$ {in linear order in the small wavevector }$\mathbf{q}${,
so that the derivation of the diffusion operator for the
isospin-singlet components would be incorrect }if coupling to the
gapful modes were neglected. The matrix equation for each set of
four Cooperons
\begin{eqnarray*}
\mathbf{C}^{l} \equiv \left(
\begin{array}{cccc}
C_{00}^{l} & C_{0x}^{l} & C_{0y}^{l} & C_{0z}^{l} \\
C_{x0}^{l} & C_{xx}^{l} & C_{xy}^{l} & C_{xz}^{l} \\
C_{y0}^{l} & C_{yx}^{l} & C_{yy}^{l} & C_{yz}^{l} \\
C_{z0}^{l} & C_{zx}^{l} & C_{zy}^{l} & C_{zz}^{l} \\
\end{array}\right),
\end{eqnarray*}
has the form
\begin{equation}
\left(
\begin{array}{cccc}
\frac{1}{2}v^{2}\tau _{0}q^{2}+\Gamma _{0}^{l}-i\omega  &
\frac{-i}{2}vq_{x}
& \frac{-i}{2}vq_{y} & 0 \\
\frac{-i}{2}vq_{x} & \frac{1}{2}\tau _{0}^{-1} & 0 & 0 \\
\frac{-i}{2}vq_{y} & 0 & \frac{1}{2}\tau _{0}^{-1} & 0 \\
0 & 0 & 0 & \tau _{0}^{-1}%
\end{array}%
\right) \mathbf{C}^{l}=1.
\end{equation}%
After the isospin-triplet modes are eliminated, the diffusion
operator for each of the four gapless/low-gap modes $C_{0}^{l}$
becomes $Dq^{2}-i\omega +\Gamma _{0}^{l}$, where
$D=\frac{1}{2}v^{2}\tau _{\mathrm{tr}}=v^{2}\tau _{0}$.

Symmetry-breaking perturbations lead to relaxation gaps $\Gamma
_{0}^{l}$ in the otherwise gapless pseudo-spin-triplet components of
the isospin-singlet Cooperon $C_{0}^{l}$. All scattering mechanisms
described in equation (\ref{h1-2}) should be included in the
corresponding disorder correlator (dashed line) on the r.h.s. of the
Bethe-Salpeter equation and in the scattering rate in the
disorder-averaged $G^{R/A}$, as $\tau _{\mathrm{0}}^{-1}\rightarrow
\tau ^{-1}=\tau _{\mathrm{0}}^{-1}+\sum_{sl}\tau _{sl}^{-1}$. This
opens
relaxation gaps in all pseudospin-triplet modes, $%
C_{0}^{x},C_{0}^{y},C_{0}^{z}$, though does not generate a
relaxation of the pseudospin-singlet $C_{0}^{0}$ which is
protected by particle conservation.

The trigonal warping term ${\hat{h}}_{\mathrm{1w}}$ in the free
electron Hamiltonian equation~(\ref{h1}) breaks the
$\mathbf{p}\rightarrow -\mathbf{p}$ symmetry of the Fermi lines
within each valley \cite{kpoints}. It has been noticed
\cite{EggShaped} that the deformation of a Fermi line of 2D
electrons in GaAs/AlGaAs heterostructures in a strong in-plane
magnetic field suppresses Cooperons as soon as the deformation
violates $\mathbf{p}\rightarrow -\mathbf{p}$ symmetry. As
${\hat{h}}_{\mathrm{1w}}$ has a similar effect, it enhances the
relaxation rate of the pseudospin-triplet intravalley components
$C_{0}^{x}$ and $C_{0}^{y}$ by
\begin{equation}
\tau _{\mathrm{w}}^{-1}=2\tau _{0}\left( \epsilon ^{2}\mu /\hbar
v^{2}\right) ^{2}.  \label{tauW}
\end{equation}%
The estimated warping-induced relaxation time is rather short for
all electron densities in the samples studied in [17],
$\tau_\mathrm{w}/\tau_{\mathrm{tr}} \sim 5 - 30, \tau_{\mathrm{w}}
< \tau_{\varphi}$, which excludes any WAL determined by
intravalley Cooperon components. However, since warping has an
opposite effect on different valleys, it does not lead to
relaxation of the pseudospin-singlet $C_{0}^{0}$ or the
intervalley component of the pseudospin triplet, $C_{0}^{z}$.

Altogether, the relaxation of modes $C_{0}^{l}$ can be described
by the following combinations of rates:
\begin{equation}\label{tau_ast_mono}
\Gamma _{0}^{0}=0,\;\Gamma _{0}^{z}=2\tau
_{\mathrm{i}}^{-1},\;\Gamma
_{0}^{x}=\Gamma _{0}^{y}=\tau _{\ast }^{-1}\equiv \tau _{\mathrm{w}%
}^{-1}+ \tau_{\mathrm{z}}^{-1} + \tau _{\mathrm{i}}^{-1},
\end{equation}
where $\tau _{\mathrm{i}}^{-1}$ is the intervalley scattering rate
(here we use the $x-y$ plane isotropy of disorder, $\tau
_{sx}^{-1}=\tau _{sy}^{-1}\equiv \tau _{s\bot }^{-1}$ and  $\tau
_{xl}^{-1}=\tau _{yl}^{-1}\equiv \tau _{\bot l}^{-1}$),
\begin{equation}\label{tau_int_mono}
\tau _{\mathrm{i}}^{-1}=4\tau _{\bot \bot }^{-1}+ 2 \tau _{z\bot
}^{-1},\; \mathrm{and}\;\tau _{\mathrm{z}}^{-1}=4\tau _{\bot
z}^{-1}+ 2\tau _{zz}^{-1}. \label{iv}
\end{equation}
After we include dephasing due to an external magnetic field,
$\mathbf{B}= \mathrm{rot}\mathbf{A}$ and inelastic decoherence,
$\tau _{\varphi }^{-1}$, the equations for $C_{0}^{l}$ read
\begin{equation}
\lbrack D(i\mathbf{\nabla }+{\textstyle\frac{2e}{c\hbar
}}\mathbf{A)}^{2}+\Gamma _{0}^{l}+\tau _{\varphi }^{-1}-i\omega
]C_{0}^{l}\left( \mathbf{r},\mathbf{ r^{\prime }}\right) = \delta
\left( \mathbf{r}-\mathbf{r^{\prime }}\right) .
\end{equation}

\subsection{Weak localization magnetoresistance in monolayer graphene}\label{subsec:magneto1}
Due to the momentum-independent form of the current operator
$\mathbf{\tilde{v}=2}v\vec{\Sigma}$, the WL correction to
conductivity $\delta g$ includes two additional diagrams,
figure~\ref{fig:2}(e) and (f) besides the standard diagram shown in
figure~\ref{fig:2}(d). Each of the diagrams in figure~\ref{fig:2}(e)
and (f) produces a contribution equal to $(-\frac{1}{4})$ of that in
figure~\ref{fig:2}(d). This partial cancellation, together with a
factor of four from the vertex corrections and a factor of two from
spin degeneracy leads to
\begin{equation}
\delta g=\frac{2e^{2}D}{\pi \hbar }\!\int \!\frac{d^{2}q}{\left(
2\pi \right) ^{2}}\left(
C_{0}^{x}+C_{0}^{y}+C_{0}^{z}-C_{0}^{0}\right) . \label{Coop-WL}
\end{equation}

Using equation~(\ref{Coop-WL}), we find the $B=0$ temperature
dependent correction, $\delta \rho /\rho =-\delta g/g$, to the
graphene sheet resistance. Taking into account the double spin
degeneracy of carriers we present
\begin{equation}
\frac{\delta \rho \left( 0\right) }{\rho ^{2}}= - \frac{e^{2}}{\pi
h}\left[ \ln (1+2\frac{\tau _{\varphi }}{\tau _{\mathrm{i}}})-2\ln
\frac{\tau _{\varphi }/\tau _{\mathrm{tr}}}{1+\frac{\tau _{\varphi
}}{\tau _{\ast }}}\right] , \label{WLZeroField}
\end{equation}%
and evaluate magnetoresistance, $\rho (B)-\rho (0)\equiv $ $\Delta
\rho (B)$,
\begin{eqnarray}
\Delta \rho (B)=\frac{e^{2}\rho ^{2}}{\pi h}\left[
F(\frac{B}{B_{\varphi }} )-F(\frac{B}{B_{\varphi
}+2B_{\mathrm{i}}})\right.  \left. -2F(\frac{B}{B_{\varphi
}+B_{\ast }})\right] ,  \label{Dsigma} \\
F(z)=\ln z+\psi (\frac{1}{2}+\frac{1}{z}),\;B_{\varphi
,\mathrm{i},\ast }= \frac{\hbar c}{4De}\tau _{\varphi
,\mathrm{i},\ast }^{-1}\,.  \nonumber
\end{eqnarray}%
Here, $\psi $ is the digamma function and the decoherence (taken
into account by the rate $\tau _{\varphi }^{-1}$) determines the
curvature of the magnetoresistance at $B < B_{\varphi} \equiv
\hbar c/4De\tau_{\varphi}$.

The last term in equation (\ref{Coop-WL}), $C_{0}^{0}$ is the only
true gapless Cooperon mode which determines the dominance of the WL
sign in the quantum correction to the conductivity in graphene with
a long phase coherence time, $\tau_{\varphi } > \tau_{\mathrm{i}}$.
The two curves sketched in figure \ref{fig:1} illustrate the
corresponding MR in two limits: $B_{\ast }\ll B_{\mathrm{i}}$ ($\tau
_{\ast }\gg \tau _{\mathrm{i}}$) and $B_{\ast }\gg B_{\mathrm{i}}$\
($\tau _{\ast }\ll \tau _{\mathrm{i}}$). In both cases, the
low-field MR ($B\ll B_{\mathrm{i}}$) is negative. If $B_{\ast }\ll
B_{\mathrm{i}}$, the MR changes sign: $\Delta \rho (B)<0$ at
$B<B_{\mathrm{i}}\equiv \hbar c/4De\tau _{\mathrm{i}}$ and $\Delta
\rho (B)>0$ at higher fields. For $B_{\ast }\gg B_{\mathrm{i}}$, the
MR is distinctly of a WL type, with almost no sign of WAL. Such
behavior is expected in graphene tightly coupled to the insulating
substrate (which generates atomically sharp scatterers). In a sheet
loosely attached to a substrate (or suspended), the intervalley
scattering time may be longer than the decoherence time, $\tau
_{\mathrm{i}}>\tau _{\varphi }>\tau _{\mathrm{w}}$
($B_{\mathrm{i}}<B_{\varphi }<B_{\mathrm{\ast }}$). Hence
$C_{0}^{z}$ is effectively gapless, whereas trigonal warping
suppresses the modes $C_{0}^{x}$ and $C_{0}^{y}$. In this case the
contribution from $C_{0}^{z}$ cancels $C_{0}^{0}$, and the MR would
display neither WL nor WAL behavior: $\Delta \rho (B)=0$.

\section{Weak localisation magnetoresistance in disordered bilayer graphene}\label{secMRbi}
\subsection{Low energy Hamiltonian of bilayer
graphene}\label{subsec:h2}
\begin{figure}
\begin{center}\resizebox{0.5\columnwidth}{!}{\includegraphics{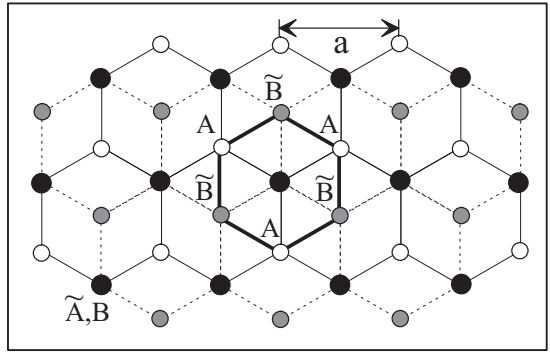}
} \caption{Schematic of the bilayer lattice (bonds in the bottom
layer $A,B$ are indicated by solid lines and in the top layer
$\tilde{A},\tilde{B}$ by dashed lines) containing four sites in
the unit cell: $A$ (white circles), $\tilde{B}$ (grey),
$\tilde{A}B$ dimer (black).} \label{fig:lattice2}
\end{center}
\end{figure}

Bilayer graphene consists of two coupled monolayers. Its unit cell
contains four inequivalent sites, $A, B, \tilde{A}$ and $\tilde{B}$
($A,B$ and $\tilde{A},\tilde{B}$ lie in the bottom and top layer,
respectively) arranged according to Bernal stacking
\cite{dressel02,McCann}: sites $B$ of the honeycomb lattice in the
bottom layer lie exactly below $\tilde{A}$ of the top layer,
figure~\ref{fig:lattice2}. The Brillouin zone of the bilayer,
similarly to the one in monolayer, has two inequivalent degeneracy
points $\mathbf{K}_+$ and $\mathbf{K}_-$ which determine two valleys
centered around $\epsilon = 0$ in the electron spectrum
\cite{kpoints}. Near the center of each valley the electron spectrum
consists of four branches. Two branches describing states on
sublattices $\tilde{A}$ and $B$ are split from energy $\epsilon = 0$
by about $\pm \gamma_1$, the interlayer coupling, whereas two
low-energy branches are formed by states based upon sublattices $A$
and $\tilde{B}$. The latter can be described \cite{McCann} using the
Hamiltonian, which acts in the space of four-component wave
functions $\Phi = [\phi_{\mathbf{K}_+,A},
\phi_{\mathbf{K}_+,\tilde{B}}, \phi_{\mathbf{K}_-,\tilde{B}},
\phi_{\mathbf{K}_-,A}]$, where $\phi_{\xi, \alpha}$ is an electron
amplitude on the sublattice $\alpha=A, \tilde{B}$ and in the valley
$\xi=\mathbf{K_+}, \mathbf{K_-}$.
\begin{eqnarray}  \label{h}
&& \!\!\!\!\!\!\!\!\!\! {\hat{H}}_{2L} = - \frac{1}{2m} \left[
\left( {p}_{x}^{2}-{p}_{y}^{2}\right)\sigma _{x} + 2 {p}_{x}{p}
_{y} \sigma _{y}
\right] + {\hat{h}}_{\mathrm{2w}} + \hat{V}_{disorder}, \\
&& {\hat{h}}_{\mathrm{2w}} = v_3 \Pi _{z}\left(p_{x}\sigma _{x} -
p_{y} \sigma _{y}\right).  \nonumber
\end{eqnarray}
Here, $\sigma_{x,y,z}$ and $\Pi_{x,y,z}$ are Pauli matrices acting
in sublattice and valley space, respectively.

The first term in equation~(\ref{h}) is the leading contribution in
the nearest neighbors approximation of the tight binding model
\cite{McCann}. This approximation takes into account both intralayer
hopping $A \leftrightarrow B $ and $\tilde{A}\leftrightarrow
\tilde{B}$ (that leads to the Dirac-type dispersion $\epsilon =\pm
pv$ near the Fermi point $\mathbf{K_{\pm }}$ in a monolayer) and the
interlayer $\tilde{A}\leftrightarrow B$ hopping. This term yields
the parabolic spectrum $\epsilon =\pm p^{2}/2m$ with
$m=\gamma_{1}/2v^{2}$ which dominates in the intermediate energy
range $\frac{1}{4}\gamma_{1}(v_{3}/v)^{2}<\varepsilon
_{F}<\frac{1}{4}\gamma _{1}$. In this regime we can truncate the
expansion of $\hat{H}(\mathbf{p})$ in powers of the momentum
$\mathbf{p}$ neglecting terms of the order higher than quadratic.
Electron waves characteristic for the first, quadratic, term of
${\hat{H}}_{2L}$ have the form
\begin{equation}
\Phi _{\mathbf{K},\pm \mathbf{p}}=\pm \frac{e^{\pm
i\mathbf{px}/\hbar }}{\sqrt{2}}\left( e^{-i\varphi }|\uparrow
\rangle_{\mathbf{K},\pm \mathbf{p}}-e^{i\varphi }|\downarrow
\rangle_{\mathbf{K},\pm \mathbf{p}}\right),
\end{equation}
where ${|\uparrow \rangle_{\mathbf{K}_{+},\pm
\mathbf{p}}}=[1,0,0,0]$, ${|\downarrow \rangle_{\mathbf{K}_{+},\pm
\mathbf{p}}}=[0,1,0,0]$ and ${|\uparrow
\rangle_{\mathbf{K}_{-},\pm \mathbf{p}}}=[0,0,1,0]$, ${|\uparrow
\rangle_{\mathbf{K}_{-},\pm \mathbf{p}}}=[0,0,0,1]$. These are
eigenstates of an operator $\mathbf{\sigma n}_{2}$ with
$\mathbf{\sigma n}_{2}=-1$ for electrons in the conduction band
and $\mathbf{\sigma n}_{2}=1$ for electrons in the valence band,
where $\mathbf{n}_{2}(\mathbf{p})=(\textrm{cos}(2\varphi
),\textrm{sin}(2\varphi ))$ for $\mathbf{p}=(p\textrm{cos}\varphi
,p\textrm{sin}\varphi )$, which means that they are chiral, but
with the degree of chirality different from the one found in
monolayer (see Sec.~\ref{subsec:h1}). Such electron waves are
characterized by the Berry phase $2\pi$, and the dependence
$w(\theta )\sim \cos ^{2}\theta $ of the scattering probability
off a short-range potential on the scattering angle $\theta =%
\widehat{\mathbf{pp}}\mathbf{^{\prime }}$ is such that transport and
scattering times in the bilayer coincide, although $w(\theta)$ is
anisotropic [see figure~\ref{fig:monopath}(c)], and the Drude
conductivity of a bilayer is $g=4e^{2}n\tau _{0}/m$ (in contrast to
monolayer graphene, see Sec.~\ref{subsec:h1}).

The second term in equation~(\ref{h}), ${\hat{h}}_{2\mathrm{w}}$,
originates from a weak direct $A \leftrightarrow \tilde{B}$
interlayer coupling. It leads to a Lifshitz transition in the shape
of the Fermi line of the 2D electron gas which takes place when
$\epsilon _{\mathrm{F}}\sim \epsilon_{\mathrm{L}}\equiv
\frac{1}{4}\gamma _{1}(v_{3}/v)^{2}$. In a bilayer with $\epsilon
_{\mathrm{F}}<\epsilon _{\mathrm{L}}$, the interplay between the two
terms in ${\hat{H}}_{2L}$ determines the Fermi line in the form of
four pockets \cite{McCann} in each valley. In a bilayer with
$\epsilon _{\mathrm{F}}>\epsilon_{\mathrm{L}}$,
${\hat{h}}_{\mathrm{2w}}$ can be treated as a perturbation leading
to a trigonal deformation of a single-connected Fermi line, thus
manifesting the asymmetry of the electron dispersion inside each
valley: $\epsilon (\mathbf{K}_{\pm },\mathbf{p})\neq \epsilon
(\mathbf{K}_{\pm },-\mathbf{p})$. This asymmetry leads to the
dephasing effect of electron trajectories similar to the one
discussed in the case of monolayer, and is characterized by the
scattering rate $\tau _{\mathrm{w}}^{-1}$ equation~(\ref{tauw1}).

The term $\hat{V}_{disorder}$ in the equation~(\ref{h}) describes
time-reversal-symmetric disorder. It is parameterized using $t
\rightarrow -t$ symmetric $4 \times 4$ matrices acting in the
sublattice/valley space, which are listed in Table~3.
\begin{eqnarray}\label{dis}
\hat{V}_{disorder} = \sum_{s,l = 0,x,y} \Pi_l \sigma_s
u_{sl}(\mathbf{r}) + \Pi_z\sigma_zu_{zz}(\mathbf{r}).
\end{eqnarray}
The sum in equation~(\ref{dis}) contains valley and isospin
conserving disorder potential $\hat{I}u(\mathbf{r})$, with $\langle
u(\mathbf{r}) u(\mathbf{r'}) \rangle = u^2
\delta(\mathbf{r}-\mathbf{r'})$ and $\tau^{-1}_0 = \pi \gamma u^2 /
\hbar, \gamma = {\textstyle\frac{m}{2\pi}} $, which originates from
charged impurities in the $\textrm{SiO}_2$ substrate and is assumed
to be the dominant mechanism of scattering in the system. All other
types of disorder which breaks valley and sublattice symmetries are
assumed to be uncorrelated, $\langle u_{s\,l}(\mathbf{r})
u_{s'\,l'}(\mathbf{r'}) \rangle = u_{s\,l}^2
\delta_{ss'}\delta_{ll'} \delta(\mathbf{r}-\mathbf{r'})$. We
characterize them using scattering rates $\tau^{-1}_{s\,l} = \pi
\gamma u^2_{s\,l}/\hbar$. Furthermore, the scattering is assumed to
be isotropic in the $x-y$ plane, so that $u^2_{xl}=u^2_{yl}\equiv
u^2_{\perp l}, u^2_{s\,x}=u^2_{sy}\equiv u^2_{s\perp}$.

\begin{table}[tbp]
\caption{ Transformations of matrices of the form
${\Pi}_l{\sigma}_s, s,l=0,x,y,z$, under crystalline symmetry group
generators and time reversal operation. In bilayer graphene
rotations and reflection symmetry operators are multiplied by the
operation of reflection with respect to the $z=0$ plane, which is
equidistant with respect to two honeycomb lattice layers. Therefore
symmetry group generators are ${\textstyle\frac{\pi}{3}}$-rotation,
$C_6R_z$, mirror reflection with respect to $Ox$ axis, $R_xR_z$, and
translation along $Ox$ by lattice constant, $\mathbf{a}$. Operation
$T$ stands for the time reversal, $t \rightarrow -t$. Here matrices
are grouped into bases forming irreducible representations of the
symmetry group which can be $1, 2$ and 4 dimensional. The
transformation matrixes $U_{ji}$, $g \phi_i = \sum_{j}
U_{ji}(g)\phi_j$ ($g$ stands for a symmetry operation), are given
explicitly for each of such bases $\phi_i$.}
\begin{center}
\begin{tabular}{|c|c|c|c|c|}
\hline\hline  $\Pi_{l}\sigma _{s}$ & $T$ &
$C_{6}R_z$ & $R_{x}R_z$ &  $\mathbf{a}$  \\
\hline $\hat{I}$ & $+1$ & $+1$ & $+1$ & $+1$  \\
\hline
$\Pi_0\sigma_{z}$ & $-1$ & $+1$ & $-1$ & $+1
$  \\
$\Pi_{z}\sigma_0$ & $-1$ & $-1$ & $+1$ & $+1$ \\
$\Pi_{z}\sigma _{z}$ & $+1$ & $-1$ &
$-1$ & $+1$  \\
$\left[\begin{array}{cc} \Pi_0\sigma_x & \\ \Pi_0\sigma_y & \\
\end{array}\right]$ &  $+1$ &
$\left(\begin{array}{cc}
  -\frac{1}{2} & \frac{\sqrt{3}}{2} \\
  -\frac{\sqrt{3}}{2} & -\frac{1}{2} \\
  \end{array}\right)$ & $\left(
\begin{array}{cc}
  1 & 0 \\
  0 & -1 \\
  \end{array}
  \right)$ &  $\left(
\begin{array}{cc}
  1 & 0 \\
  0 & 1 \\
  \end{array}
  \right)$  \\
$ \left[
\begin{array}{cc}
  \Pi_z\sigma_x  \\
  \Pi_z\sigma_y   \\
  \end{array}
\right]$ & $-1$ & $\left(
\begin{array}{cc}
   \frac{1}{2} & -\frac{\sqrt{3}}{2} \\
  \frac{\sqrt{3}}{2} & \frac{1}{2} \\
  \end{array}
  \right)$ & $\left(
\begin{array}{cc}
  1 & 0 \\
  0 & -1 \\
  \end{array}
  \right)$ & $\left(
\begin{array}{cc}
  1 & 0 \\
  0 & 1 \\
  \end{array}
  \right)$   \\
$\left[
\begin{array}{cc}
  \Pi_x\sigma_0  \\
  \Pi_y\sigma_0   \\
  \end{array}
\right]$ & $+1$ & $\left(
\begin{array}{cc}
  1 & 0 \\
  0 & -1 \\
  \end{array}
  \right)$ & $\left(
\begin{array}{cc}
  1 & 0 \\
  0 & 1 \\
  \end{array}
  \right)$ &  $\left(
\begin{array}{cc}
  -\frac{1}{2} & -\frac{\sqrt{3}}{2} \\
  \frac{\sqrt{3}}{2} & -\frac{1}{2} \\
  \end{array}
  \right)$ \\
$\left[
\begin{array}{cc}
  \Pi_x\sigma_z  \\
  \Pi_y\sigma_z   \\
  \end{array}
\right]$ & $-1$ & $\left(
\begin{array}{cc}
  1 & 0 \\
  0 & -1 \\
  \end{array}
  \right) $ & $\left(
\begin{array}{cc}
  1 & 0 \\
  0 & 1 \\
  \end{array}
  \right)$ & $\left(
\begin{array}{cc}
   -\frac{1}{2} & -\frac{\sqrt{3}}{2} \\
  \frac{\sqrt{3}}{2} & -\frac{1}{2} \\
  \end{array}
  \right)$  \\
$\left[
\begin{array}{cccc}
  \Pi_x\sigma_x  \\
  \Pi_x\sigma_y  \\
  \Pi_y\sigma_x   \\
  \Pi_y\sigma_y   \\
  \end{array}
\nonumber \right]$ & $+1$ & $\left(
\begin{array}{cccc}
  -\frac{1}{2} & \frac{\sqrt{3}}{2} & 0 & 0 \\
  -\frac{\sqrt{3}}{2} & -\frac{1}{2} & 0 & 0 \\
  0 & 0 & \frac{1}{2} & -\frac{\sqrt{3}}{2} \\
  0 & 0 &  \frac{\sqrt{3}}{2} & \frac{1}{2} \\
  \end{array}
  \right)$ & $\left(
\begin{array}{cccc}
  1 & 0 & 0 & 0 \\
  0 & -1 & 0 & 0 \\
  0 & 0 & 1 & 0 \\
  0 & 0 & 0 & -1 \\
  \end{array}
  \right)$ & $\left(
\begin{array}{cccc}
  -\frac{1}{2} & 0 & -\frac{\sqrt{3}}{2} & 0 \\
  0 &  -\frac{1}{2} & 0 & -\frac{\sqrt{3}}{2} \\
  \frac{\sqrt{3}}{2} & 0 & -\frac{1}{2} & 0 \\
  0 & \frac{\sqrt{3}}{2} & 0 & -\frac{1}{2} \\
  \end{array}
  \right)$  \\
\hline\hline
\end{tabular}
\end{center}
\end{table}

%
\subsection{Interference of electronic waves in bilayer graphene}\label{subsec:qual2}
To analyze the WL effect we introduce Cooperon matrix $C_{\alpha
\beta \alpha ^{\prime }\beta ^{\prime }}^{\xi \mu \xi ^{\prime
}\mu ^{\prime }}$ where subscripts describe the sublattice state
of incoming $\alpha \beta $ and outgoing $\alpha ^{\prime }\beta
^{\prime }$ pairs of electrons and superscripts describe the
valley state of incoming $\xi \mu $ and outgoing $\xi ^{\prime
}\mu ^{\prime }$ pairs. Note that in contrast to monolayer we do
not rewrite the bilayer Hamiltonian in terms of $\Sigma$ and
$\Lambda$ matrices. We parametrize Cooperons as
$C_{S_{1}S_{2}}^{M_{1}M_{2}}$ by $M_1, M_2$ "valley" and $S_1,S_2$
"sublattice" singlet and triplet states in a similar way to
monolayer isospin and pseudospin states. The sublattice
composition of Cooperons is determined by the correlator of plane
waves propagating ballistically in opposite directions,
\begin{eqnarray*}
\Phi _{\mathbf{K},\mathbf{p}}\Phi
_{\mathbf{K}^{\prime},-\mathbf{p}} \!\sim \! {|\uparrow \rangle
}_{\mathbf{K},\mathbf{p}}|\downarrow \rangle
_{\mathbf{K}^{\prime},-\mathbf{p}}\! +\!|\downarrow \rangle
_{\mathbf{K},\mathbf{p}}|\uparrow \rangle_{\mathbf{K}^{\prime},
-\mathbf{p}}\! -e^{2i\varphi }|\uparrow\rangle
_{\mathbf{K},\mathbf{p}}|\uparrow \rangle
_{\mathbf{K}^{\prime},-\mathbf{p}}\!-e^{-2i\varphi }|\downarrow
\rangle _{\mathbf{K},\mathbf{p}}|\downarrow \rangle
_{\mathbf{K}^{\prime},-\mathbf{p}}.
\end{eqnarray*}
It is seen from the above expression that after averaging over the
momentum direction the terms corresponding to $C_{x,y}^{M}\propto
({|\uparrow \rangle_{\mathbf{K},\mathbf{p}}}{|\uparrow \rangle
_{\mathbf{K^{\prime }}, -\mathbf{p}}}\pm {|\downarrow \rangle
_{\mathbf{K},\mathbf{p}}}{|\downarrow \rangle_{\mathbf{K^{\prime
}},-\mathbf{p}}})$ disappear, since $\mathbf{p}=(p\cos\varphi,
p\sin\varphi)$ so that $\langle e^{\pm 2i\varphi }\rangle
_{\varphi }=0$, whereas terms correponding to the sublattice
symmetric Cooperons, $C_{z}^{M}\propto ({|\uparrow \rangle
_{\mathbf{K},\mathbf{p}}}{|\downarrow \rangle _{\mathbf{K^{\prime
}},-\mathbf{p}}}+{|\downarrow \rangle
_{\mathbf{K},\mathbf{p}}}{|\uparrow \rangle _{\mathbf{K^{\prime
}},-\mathbf{p}}})$ remain non-zero.

The dephasing effect of trigonal warping in bilayer is similar to
monolayer, although it is caused by a different mechanism, its
magnitude is estimated by equation~(\ref{tauw1}). Dephasing due to
warping suppresses the intravalley Cooperons $C_{z}^{x,y}$ leading
to the absence of WL magnetoresistance in the case of weak
intervalley scattering, $\tau_{\mathrm{i}} \gg \tau_\varphi$. In the
case of strong intervalley scattering, $\tau_{\mathrm{i}} \ll
\tau_\varphi$, WL is partially restored, thus, we predict the WL
behavior of bilayer graphene with strong trigonal warping of Fermi
line in each valley to be described by equation~(\ref{summary}).

\subsection{Diagrammatic calculation of the weak localization
correction in bilayer graphene}\label{subsec:diagram2}
We derive the disorder averaged Green function for the bilayer
Hamiltonian equation~(\ref{h}):
\begin{equation}
G^{R/A}\left( \mathbf{p},\epsilon \right) = \frac{\epsilon_{R/A} -
\epsilon_{p}\,\mathbf{\sigma
n_{2}}(\mathbf{p})}{\epsilon_{R/A}^{2} - \epsilon _{p}^{2}},
\end{equation}
where $\epsilon _{R/A}=\epsilon \pm {\textstyle\frac{1}{2}}i\hbar
\tau ^{-1}$ and $\tau ^{-1}=\tau _{0}^{-1} + \tau_{\mathrm{i}}^{-1}
+\tau _{\mathrm{z}}^{-1}\approx \tau_{0}^{-1}$. Here we introduced
the following notations for the scattering rates $\tau
_{sl}^{-1}=\pi \gamma u_{sl}^{2}/\hbar$, where $\tau _{sx}^{-1}=\tau
_{sy}^{-1}\equiv \tau _{s\bot }^{-1}$ and $\tau _{xl}^{-1}=\tau
_{yl}^{-1}\equiv \tau _{\bot l}^{-1}$ can be combined into the
intervalley scattering rate $\tau_{\mathrm{i}}^{-1}=4\tau _{\bot
\bot }^{-1}+2\tau _{z\bot }^{-1}$ and the intravalley rate
$\tau_{\mathrm{z}}^{-1}=\tau _{zz}^{-1}$ both of which lead to an
additional suppression of intravalley modes. Intervalley scattering
also leads to the relaxation of $C_{z}^{0}$ although it does not
affect the valley-symmetric mode $C_{z}^{z}$. Together, all the
scattering mechanisms limit the transport time $\tau^{-1} =
\tau_{0}^{-1} + \sum_{sl} \tau_{sl}^{-1}$.

Due to quadratic spectrum of quasiparticles in bilayer graphene the
velocity operator, $\hat{v}_{x}=-( p_{x}\sigma_{x} +
p_{y}\sigma_{y})/m $, $\hat{v}_{y}=( p_{y}\sigma_{x} -
p_{x}\sigma_{y})/m$, is momentum dependent, and thus the current
vertices in the conductivity diagram figure~\ref{fig:2}(c) are not
renormalized by impurity scattering accounted for by the diagram
series figure~\ref{fig:2}(b). As a result, the Drude conductivity is
described by $g = 4e^2 \nu D$, where $D={\textstyle\frac{1}{2}}v^2_F
\tau_0$ and $\tau_{tr} = \tau_0$.

We parametrize the Cooperons utilizing the expression,
\begin{eqnarray*}
C^{M_1M_2}_{S_{1}S_{2}} &=& \frac{1}{4} \sum_{\alpha, \beta,
\alpha', \beta'} \sum_{\xi,\mu,\xi',\mu'} \left( \sigma_y
\sigma_{S_1} \right)_{\alpha\beta} \left( \Pi_y\Pi_{\kappa_1}
\right)^{\xi\mu} C^{\;\xi\mu,\xi'\mu}_{\alpha\beta,\alpha'\beta'}
\left( \sigma_{S_2} \sigma_y \right)_{\beta'\alpha'} \left(
\Pi_{\kappa_2}\Pi_y \right)^{\mu'\xi'}.
\end{eqnarray*}
The Bethe-Salpeter equation for Cooperons in bilayer reads,
\begin{eqnarray}\label{bs1}
C_{S_{1}S_{2}}^{M_{1}M_{2}}\left( \mathbf{q}\right) &=&\tau
_{0}\,\delta _{M_{1},M_{2}}\delta _{S_{1},S_{2}}  \nonumber \\
& \!\!\!\!\!\!\!\!\!\!\!\!\!\!\!\!\!\!\!\!\!\!\!\!\!\!\!\!\!\!\!\!\!\!\!\!\!\!\!\!\!\!\!\!\!\!\!\!\!\!\!\!\!+&\!\!\!\!\!\!\!\!\!\!\!\!\!\!\!\!\!\!\!\!\!\!\!\!\!\!\!\frac{1}{4\pi \gamma \tau _{0}\hbar }%
\sum_{S,M}C_{SS_{2}}^{MM_{2}}\left( \mathbf{q}\right) \int
\frac{d^{2}p}{\left( 2\pi \right) ^{2}} \mathrm{Tr}\Big\{\Pi
_{M}\sigma _{S}\Pi_{y}\sigma _{y}\left[ G_{\mathbf{p},\hbar \omega
+\epsilon }^{R}\right] ^{\mathrm{t}}\Pi _{y}\sigma _{y}\Pi
_{M_{1}}\sigma _{S_{1}}G_{\hbar \mathbf{q}-\mathbf{p},\epsilon
}^{A}\Big\} .
\end{eqnarray}
We find that $C_{SS^{\prime }}^{MM^{\prime }}=\delta ^{MM^{\prime
}}\delta _{SS^{\prime }}C_{S}^{M}$ and that sublattice-singlet
$C_{0}^{M}$ has a relaxation gap $\Gamma _{0}^{M}=\tau _{0}^{-1}$,
sublattice-triplets $C_{x}^{M}$, $C_{y}^{M}$ have gaps $\Gamma
_{x}^{M}=\Gamma _{y}^{M}=\frac{1}{2}\tau _{0}^{-1}$, whereas
symmetric sublattice-triplet Cooperon $C_{z}^{M}$ is gapless. Due
to warping of the Fermi line induced by ${\hat{h}_{2w}}$ in the
free-electron Hamiltonian (\ref{h}), the intravalley Cooperons
$C_{z}^{x}$, $C_{z}^{y}$ are suppressed, even in a bilayer with
purely potential disorder. Warping opens a gap, $\tau
_{\mathrm{w}}^{-1}$ in the relaxation spectrum of these
`valley-triplet' Cooperon components:
\begin{equation}
\tau_{\mathrm{w}}^{-1} = \left\{
\begin{array}{cc}
{\textstyle\frac{1}{2\hbar^2}} \tau \langle
\mathrm{Tr}{\hat{h}}_{\mathrm{w}}^{2}(\mathbf{p})\rangle_{\varphi
} = \pi n_L l^2 \tau^{-1}, & \pi n_L l^2 < 1 \\
\!\!\!\!\!\!\!\!\!\!\!\!\!\!\!\!\!\!\!\!\!\!\!\!\!\!\!\!\!\!\!\!\!\!\!\!\!\!\!\!\!\!\!\!\!\!\!\!\!\!\!\!\!\!\!\!\!
\tau^{-1}, & \pi n_L l^2 > 1 \\
\end{array}\right., \label{tauw}
\end{equation}
where $n_L$ is the density of electrons at which Lifshitz
transition occurs, $l$ and $\tau$ are mean free path and transport
time in the system respectively. We estimate that for the recently
studied bilayers \cite{QHEbi} with $n_{e}= 2.5 \times
10^{12}\textrm{cm}^{-2}$, $\tau_{\mathrm{w}} \sim \tau$ and the
mean free path $l \sim 0.1 \mu\textrm{m}$. A similar situation
occurs in bilayer structures studied by R.~Gorbachev {\em et al.}
\cite{exeter}.
Also, a short-range symmetry-breaking disorder $u_{ij}$ generating
intervalley scattering leads to the relaxation of $C_{z}^{0}$,
although it does not affect the valley-symmetric mode $C_{z}^{z}$.
Thus we find that the low-gap modes $C_{z}^{M}$ obey the diffusion
equation,
\begin{eqnarray}\label{lowgap}
&&\!\!\!\!\!\!\!\!\left[ \Gamma +\tau _{\varphi }^{-1}+
D(i\mathbf{\nabla }+{\textstyle\frac{2e}{c\hbar
}}\mathbf{A)}^{2}-i\omega \right] C\left(
\mathbf{r},\mathbf{r^{\prime }}\right) =
\delta \left( \mathbf{r}-\mathbf{r^{\prime }}\right) ,  \nonumber \\
&&\!\!\!\!\!\!\!\!\Gamma _{z}^{z}=0,\;\Gamma _{z}^{0}=2\tau _{\mathrm{i}}^{-1},\; \\
&&\!\!\!\!\!\!\!\!\Gamma _{z}^{x(y)}=\tau _{\ast }^{-1}\equiv \tau
_{\mathrm{w}}^{-1}+2\tau _{\mathrm{z}}^{-1}+\tau
_{\mathrm{i}}^{-1}, \nonumber
\end{eqnarray}%
where we included dephasing due to an external magnetic field,
$\mathbf{B}=\mathrm{rot}\mathbf{A}$, temperature-dependent
inelastic decoherence, $\tau _{\varphi }^{-1}(T)$, and all of the
above mentioned relaxation mechanisms.

\subsection{Weak localization magnetoresistance in bilayer graphene}\label{subsec:magneto2}
The interference correction to the conductivity in a
bilayer can be expressed in terms of $C\left(
\mathbf{r},\mathbf{r}\right) $, the solutions of the above
Cooperon equations taken at coinciding coordinates:
\begin{eqnarray}\label{wlc}
\delta g=\frac{2e^{2}D}{\pi \hbar }\left[ - C_{z}^{z} + C_{z}^{0}
- C_{z}^{x} - C_{z}^{y} \right].
\end{eqnarray}
For completeness, in equation~(\ref{wlc}) we have retained the
intravalley Cooperons $C_{z}^{x,y}$, though they are strongly
suppressed by trigonal warping. Following their suppression, the WL
correction is determined by the intervalley modes $C_{z}^{0}$ and
$C_{z}^{z}$ but, in the absence of intervalley scattering, the
contributions of $C_{z}^{0}$ and $C_{z}^{z}$ are equal in magnitude,
so that they cancel. Intervalley scattering due to atomically sharp
scatterers breaks this exact cancellation and partially restores the
WL effect. Equations (\ref{wlc},\ref{lowgap}) yield the zero field
WL correction to the resistivity and the WL MR,
\begin{eqnarray}
\frac{\delta \rho \left( 0\right) }{\rho} &=& \frac{e^{2}\rho}{\pi
h} \ln \left( 1 + 2\frac{\tau_{\varphi }}{\tau_{\mathrm{i}}}
\right) + \delta_0 , \label{MRbi} \\
\frac{\Delta \rho (B)}{\rho} &=& -\frac{e^{2}\rho}{\pi h} \left[
F(\frac{B}{B_{\varphi }}) - F(\frac{B}{B_{\varphi }+
2B_{\mathrm{i}}}) \right] + \delta (B) , \nonumber
\end{eqnarray}
where $B_{\varphi ,\mathrm{i}}= \hbar c/(4De\tau_{\varphi,
\mathrm{i}})$. Equation~(\ref{MRbi}) gives a complete description of
the crossover between two extreme regimes mentioned at the beginning
\cite{socomment}. It also includes small contributions of the
suppressed intravalley Cooperons, $\delta_0 = [2 e^{2}\rho /(\pi h)]
\ln ( \tau_{\varphi} \tau_{\ast} / [\tau (\tau_{\ast} +
\tau_{\varphi})])$ and $\delta (B) = - [2 e^{2}\rho /(\pi h)] F[ B/
(B_{\varphi} + B_{\ast})]$, where $\tau_{\ast}^{-1} =
\tau_{\mathrm{w}}^{-1} + 2\tau_{\mathrm{z}}^{-1} +
\tau_{\mathrm{i}}^{-1}$ and $B_{\ast}= \hbar c/(4De\tau_{\ast})$.
This permits us to account for a possible difference between the
warping time $\tau_{\mathrm{w}}$ and the transport time $\tau$.
According to equation~(\ref{MRbi}) WL MR in bilayer graphene sheet
disappears as soon as $\tau_{\mathrm{i}}$ exceeds $\tau_{\varphi}$,
whereas in structures with $\tau_{\varphi} > \tau_{\mathrm{i}}$, the
result equation~(\ref{MRbi}) predicts the WL behaviour, as observed
in \cite{exeter}. Such WL MR is saturated at a magnetic field
determined by the intervalley scattering time, instead of the
transport time as in usual conductors, which provides the
possibility to measure $\tau_{\mathrm{i}}$ directly.

\section{Conclusions and the effect of edges in a disordered nanoribbon}\label{secconc}
We have shown that $\mathbf{p} \rightarrow - \mathbf{p}$ asymmetry
of the electron dispersion in each valley of graphene leads to
unusual (for conventional disordered conductors) behavior of
interference effects in electronic transport. Without intervalley
scattering, trigonal warping of the electron dispersion near the
center of each valley destroys the manifestation of chirality in the
localization properties, resulting in a suppression of weak
anti-localization in monolayer graphene and of weak localization in
a bilayer. Intervalley scattering tends to restore weak
localization, and this behavior is universal for monolayer and
bilayer graphene, despite the fact that electrons in these two
materials have different chiralities and can be attributed different
Berry phases: $\pi$ in monolayers, $2\pi$ in bilayers
\cite{haldane88,McCann}. This suggests that a suppressed weak
localization magnetoresistance and its sensitivity to intervalley
scattering are specific to all ultrathin graphitic films
independently of their morphology \cite{berger} and are determined
by the lower (trigonal) symmetry group of the wavevector
$\mathbf{K}$ in the corner of the hexagonal Brillouin zone of a
honeycomb lattice crystal.

The influence of intervalley scattering on the WL behavior
determines a typically negative (WL) MR in graphene nanoribbons.
Indeed, in a narrow ribbon of graphene, monolayer or bilayer, with
the transverse diffusion time $L_{\bot }^{2}/D\ll \tau
_{\mathrm{i}},\tau _{\ast },\tau _{\varphi }$, the sample edges
determine strong intervalley scattering rate \cite{edge}. Thus,
when solving Cooperon equations in a wire, we estimate $\Gamma
_{0}^{l}\sim \pi ^{2}D/L_{\bot }^{2} $ for the pseudospin triplet,
whereas the singlet $C_{0}^{0}$ remains gapless. This yields
negative MR persistent over the field range $B < 2\pi B_{\bot }$,
where $B_{\bot }\equiv \hbar c/eL_{\bot }^{2}$:
\begin{equation}
\frac{\Delta \rho _{\mathrm{wire}}\left( B\right) }{\rho ^{2}}=\frac{%
2e^{2}L_{\varphi }}{h}\,\left[
\frac{1}{\sqrt{1+\frac{1}{3}B^{2}/B_{\varphi }B_{\bot }}}-1\right]
.  \label{WL1D}
\end{equation}

The results of Eqs.
(\ref{WLZeroField},\ref{Dsigma},\ref{wlc},\ref{MRbi}), and
(\ref{WL1D}) give a complete description of the WL effect in
graphene and describe how the WL magnetoresistance reflects the
degree of valley symmetry breaking in it.

This project has been funded by Lancaster-EPSRC Portfolio
Partnership grant EP/C511743 and was completed during the MPI PKS
Seminar "Dynamics and Relaxation in Complex Quantum and Classical
Systems and Nanostructures."


\begin{thebibliography}{}
\bibitem{socomment} In this calculation we take into account
double spin degeneracy of carriers.
\bibitem{t-rev} For the monolayer, time reversal of an operator $\hat{W}$
is described by $(\Pi _{x}\otimes \sigma _{x}){\hat{W}}^{\ast
}(\Pi _{x}\otimes \sigma _{x})$. For the bilayer, time reversal is
given by $\Pi_{x}\otimes\sigma_{x}{\hat{W}}^{\ast} \Pi_{x} \otimes
\sigma _{x}$.
\bibitem{U4} The group U$_{4}$ can be described using 16 generators
$\mathrm{\hat{I}},\Sigma _{s},\Lambda _{l},\Sigma _{s}\Lambda _{l}$,
$s,l=x,y,x$.
\bibitem{kpoints} Corners of the hexagonal Brilloin zone are $\mathbf{K}_{\pm }=\pm
({\textstyle\frac{2}{3}}ha^{-1},0)$.
\end{thebibliography}
\end{document}